\documentclass[a4paper]{spie}  

\usepackage[]{graphicx}

\newcommand{\dazle}{DA$z$LE}
\newcommand{\micron}{\ensuremath{\rm{\mu m}}}
\newcommand{\degree}{\ensuremath{^{\circ}}}

\newcommand{\lya}{\mbox{Lyman-}\ensuremath{\alpha}}
\newcommand{\ly}{\mbox{Ly}\ensuremath{\alpha}}

\newcommand{\intpa}{\ensuremath{\gamma\ \mbox{s}^{-1}\mbox{m}^{-2}\mbox{arcsec}^{-2}}}

\title{DA\textit{\LARGE z}LE: The Dark Ages \textit{\LARGE z} (redshift) Lyman-$\mbox{\LARGE $\alpha$}$ Explorer} 

\author{Anthony Horton\supit{a}, Ian Parry\supit{a}, Joss Bland-Hawthorn\supit{b}, Sonia Cianci\supit{b}, David King\supit{a}, Richard McMahon\supit{a}, Steve Medlen\supit{a}
\skiplinehalf
\supit{a}Institute of Astronomy, Madingley Road, Cambridge, UK; \\
\supit{b}Anglo-Australian Observatory, 167 Vimiera Rd, Epping, NSW, Australia
}

\authorinfo{A. Horton, E-mail: horton@ast.cam.ac.uk, Telephone: +44 1223 339279}

\begin{document} 
  \maketitle 

\begin{abstract}
\dazle\ is an near infrared narrowband differential imager being built by the Institute of Astronomy, Cambridge, in collaboration with the Anglo-Australian observatory.  It is a special purpose instrument designed with a sole aim; the detection of redshifted \lya\ emission from star forming galaxies at $z>7$.  \dazle\ will use pairs of high resolution ($R=1000$) narrowband filters to exploit low background `windows' in the near infrared sky emission spectrum.  This will enable it to reach sensitivities of $\sim2\times10^{-21}$Wm$^{-2}$, thereby allowing the detection of $z>7$ galaxies with star formation rates as low as a few solar masses per year.  The design of the instrument, and in particular the crucial narrowband filters, are presented.  The predicted performance of \dazle\, including the sensitivity, volume coverage and expected number counts, is discussed.  The current status of the \dazle\ project, and its projected timeline, are also presented.
\end{abstract}



\section{INTRODUCTION}
\label{sect:intro}

The question of how and when galaxies formed is one at the forefront of cosmology, and direct observations of the youngest galaxies possible are a vital part of advancing our understanding of the process of galaxy formation.  While the faintness of extremely distant sources makes detailed study of them difficult important information can be gleaned from the number densities, luminosities and clustering properties.

The use of optical wavelength narrowband filters has been very successful in detecting high redshift galaxies via their redshifted \lya\ emission, allowing direct observations at redshifts of up to $\sim6.6$ (see, for example, Hu et al 2002\cite{Hu2002} and Kodaira et al 2003\cite{Kodaira2003}).  The impressive results obtained so far suggest it should be possible to extend this technique to even higher redshifts, especially in light of recent data which implies that signficant star formation was underway at $z>7$ (for instance the electron scattering optical depth results from the Wilkinson Microwave Anisotripy Probe (WMAP)\cite{WMAP2003a} and FeII/MGII line ratios from high redshift SDSS quasar spectra\cite{dietrich2003,corbin2003,maiolino2003,barth2003}).  Higher reshifts will require the use of infrared instrumentation, however, such as \dazle.

\dazle\ is a specialised near infrared narrowband differential imager designed for the detection of redshifted \lya\ emission from $z > 7$.  It is being built by the Institute of Astronomy, Cambridge (IoA) in collaboration with the Anglo-Australian Observatory (AAO), and is intended for use on the visitor focus of the ESO VLT.  \dazle\ will use pairs of high resolution ($R=1000$) filters closely spaced in wavelength to select line emitters via differential imaging.  The wavelengths of the filters are chosen to fall in the very low background `windows' in the sky's OH emission line spectrum thereby giving \dazle\ the sensitivity it needs to detect faint line emission from the early universe.  The \dazle\ \lya\ surveys will target fields in which deep broadband visible and near infrared imagery is available, for instance the GOODS-Chandra Deep Field South fields.  The combined data will enable the selection of high redshift \lya\ emitters.

\section{INSTRUMENT DESIGN}
\label{sect:design}

\subsection{Overview}
\label{sect:design:overview}

\dazle\ is a near infrared re-imaging instrument capable of operation over the wavelength range 1.0--1.8\micron\ and currently optimised for the range 1.06--1.33\micron\ (which corresponds to \lya\ redshifts of $\sim7.7$--10.0).  It will use the cryogenic camera from the CIRPASS spectrograph\cite{ajd}.  The use of a pre-existing camera saves considerable time and expense.  At the heart of the CIRPASS camera is a HAWAII-2 $2048\times2048$ pixel detector.  The chosen pixel scale for \dazle\ is $0.2''$ per pixel, giving a large field of view of $6.83'\times6.83'$.

\begin{figure}[tbp]
\includegraphics[width=\textwidth]{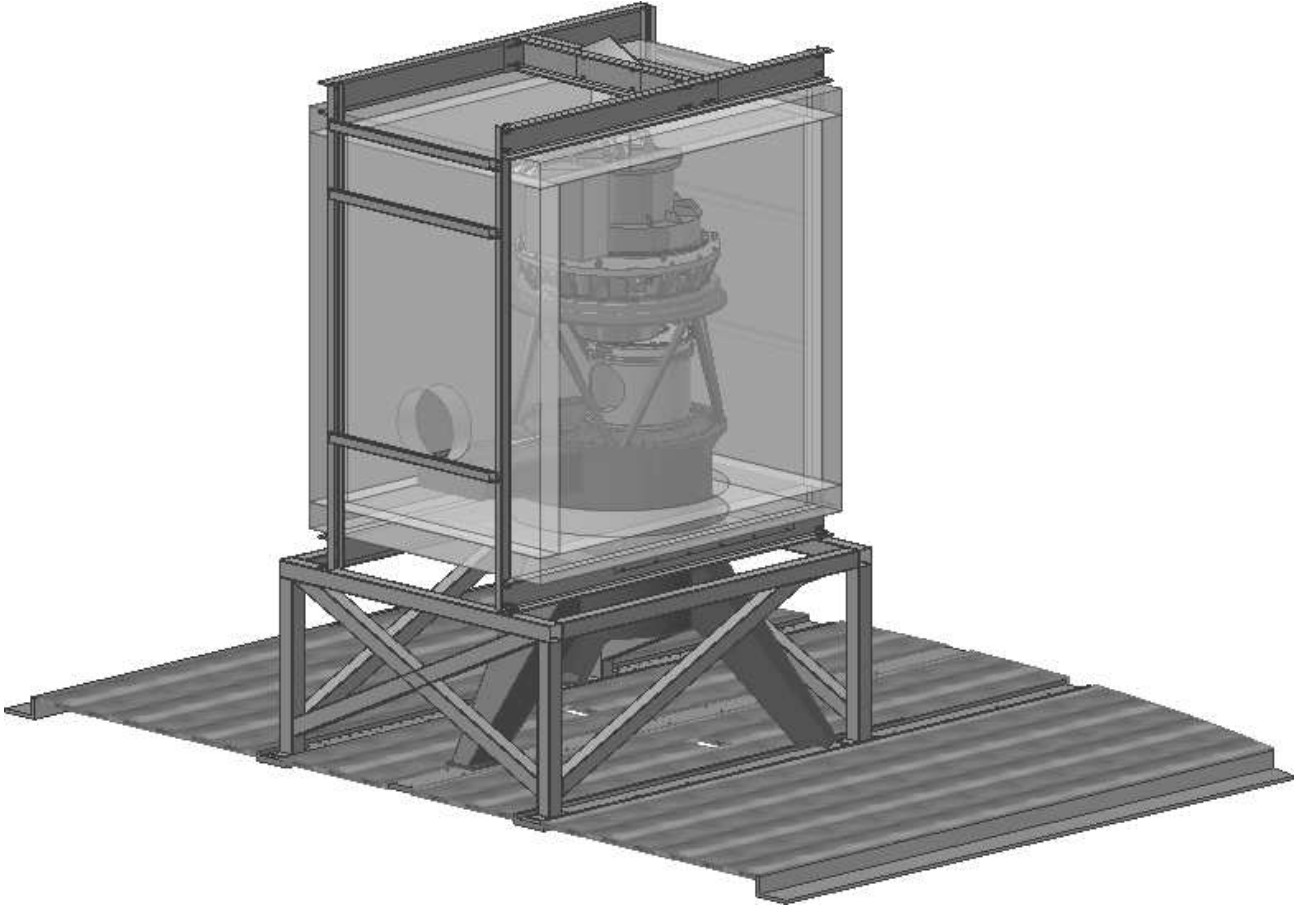}
\caption{Schematic of \dazle\ in position on the VLT Nasmyth platform.  The overall height of the instrument is approximately 4 metres.}
\label{fig:overview}
\end{figure}

Fig.\ \ref{fig:overview} is a schematic of \dazle\ in place on the VLT Nasmyth platform.  The entire instrument is contained within an insulated enclosure, shown in the figure as a semi-transparent box, and cooled to -40\degree C to reduce thermal infrared emission.  This technique has previously been used by the IoA with CIRPASS and results in considerable reductions in instrument background beyond that which can be achieved with blocking filters alone. Low instrument background is crucial for \dazle\ to take full advantage of the low sky backgrounds provided by the narrowband filters.  The light from the telescope enters the instrument through a collimator (not shown) at approximately 2 metres above the Nasmyth platform, and is then directed upwards by a fold mirror to pass through filter wheels, a cold stop, and finally into the CIRPASS camera.  The instrument as a whole will stand approximately 4 metres high.

\subsection{Narrowband Filters}
\label{sect:design:filters}

\begin{table}[tp]
\centering
\caption[The 15 widest OH windows in the J-band]{The 15 widest OH windows in the J-band.  For each window the vacuum wavelength and intensity of the OH line at each end are given, in \AA\ and \intpa\ respectively.  The width of the window is given in \AA\ and the approximate atmospheric transmission as a percentage.  The first column gives the \lya\ redshift corresponding to the centre of the window.\\}
\label{table:windows}
\begin{tabular}{|r|r|r|r|r|r|r|l|}
\hline
\rule[-1ex]{0pt}{3.5ex} $z_c$ & \multicolumn{2}{c|}{Start Line} & Width & \multicolumn{2}{c|}{End Line} & Trans. & Comments \\
\cline{2-3} \cline{5-6}
\rule[-1ex]{0pt}{3.5ex} & \multicolumn{1}{c|}{$\lambda$} & \multicolumn{1}{c|}{Int.} & & \multicolumn{1}{c|}{$\lambda$} & \multicolumn{1}{c|}{Int.} & (\%) & \\
\hline
\rule[-1ex]{0pt}{3.5ex} 7.68 & 10528.0 & 13.55 & 47.3 & 10575.2 & 2.51 & 100 & Clean window. \\
\hline
\rule[-1ex]{0pt}{3.5ex} 7.73 & 10589.2 & 4.93 & 54.3 & 10643.4 & 0.75 & 100 & Clean window. \\
\hline
\rule[-1ex]{0pt}{3.5ex} 7.88 & 10775.1 & 11.68 & 57.1 & 10832.2 & 8.98 & 100 & Clean window. \\
\hline
\rule[-1ex]{0pt}{3.5ex} 8.14 & 11090.4 & 22.23 & 50.4 & 11140.7 & 4.28 & 90 & Clean window. \\
\hline
\rule[-1ex]{0pt}{3.5ex} 8.38 & 11377.2 & 14.66 & 60.6 & 11437.8 & 11.04 & 80 & Possible contamination. \\
\hline
\rule[-1ex]{0pt}{3.5ex} 8.60 & 11651.0 & 57.73 & 45.3 & 11696.3 & 13.68 & 95 & Some contamination. \\
\hline
\rule[-1ex]{0pt}{3.5ex} 8.66 & 11716.5 & 30.11 & 54.1 & 11770.7 & 6.05 & 95 & Possible contamination. \\
\hline
\rule[-1ex]{0pt}{3.5ex} 8.72 & 11788.5 & 12.14 & 62.7 & 11851.2 & 2.07 & 95 & Clean window.\\
\hline
\rule[-1ex]{0pt}{3.5ex} 8.79 & 11867.1 & 3.85 & 71.0 & 11938.0 & 0.55 & 95 & Some contamination. \\
\hline
\rule[-1ex]{0pt}{3.5ex} 8.94 & 12055.9 & 16.90 & 64.5 & 12120.4 & 12.46 & 95 & Some contamination. \\
\hline
\rule[-1ex]{0pt}{3.5ex} 9.18 & 12351.9 & 68.99 & 48.9 & 12400.8 & 16.89 & 100 & Some contamination. \\
\hline
\rule[-1ex]{0pt}{3.5ex} 9.24 & 12423.7 & 37.60 & 58.8 & 12482.5 & 7.88 & 100 & Some contamination. \\
\hline
\rule[-1ex]{0pt}{3.5ex} 9.84 & 13157.1 & 78.19 & 53.7 & 13210.8 & 19.80 & 95 & Clean window. \\
\hline
\rule[-1ex]{0pt}{3.5ex} 9.91 & 13236.9 & 44.57 & 64.8 & 13301.8 & 9.76 & 90 & Some contamination. \\
\hline
\rule[-1ex]{0pt}{3.5ex} 9.99 & 13325.2 & 20.5 & 75.7 & 13401.0 & 3.81 & 80 & Clean window. \\
\hline
\end{tabular}
\end{table}

As mentioned in Sect. \ref{sect:intro} \dazle\ gets the sensitivity it needs by exploiting low background windows in the sky emission spectrum with very narrowband filters.  The first step in the design of the \dazle\ filters was therefore to indentify the best windows to use.  The majority of the sky background is in the form of emision lines from OH radicals in the upper atmosphere.  To find the widest gaps between OH lines we used the theoretical line list prepared by Rousselot et al\cite{Rousselot_et_al}, which is based on the energy levels published by Abrams et al\cite{Abrams} and the transition probabilities due to Mies\cite{Mies}.  The Rousselot et al line list gives only relative intensities, and so before use it was calibrated by comparision to the observed OH line intensities of Maihara et al\cite{Maihara}.

At this point it is possible to compile a list of the widest gaps in the OH emission spectrum.  The OH lines are not the only emission lines in the sky spectrum however, and so the gaps were checked for O$_2$ lines or other contamination by inspecting an ISAAC observed sky spectrum, also published by Rousselot et al\cite{Rousselot_et_al}.  In addition the atmospheric transmission for the gap wavelengths was obtained from data published by the Gemini Observatory.  This data was generated by the ATRAN modelling software due to Lord\cite{Lord}.  The remaining J-band windows not compromised by O$_2$ emission or atmospheric absorption are listed in Table \ref{table:windows}.

In order to specify the filters for \dazle\ it was then necessary to select windows from this list.  The \dazle\ differential imaging concept calls for pairs of filters, ideally as close together in wavelength as possible to give good subtraction of sky background and continuum sources.  It seems prudent in the first instance to target relatively low redshifts.  The first two windows in the J-band, at $\sim1.06$\micron, are ideal for the first \dazle\ filter pair.  Their central redshifts of 7.68 and 7.73 do not represent an unreasonable leap up from the current \lya\ detections at $z\sim6.6$ however the difference is sufficient to be scientifically interesting.  To continue the experiment to higher redshifts a second set of filters will be needed.  The next set of closely spaced windows are the four clustered around 1.18\micron.  The two cleanest windows, centred on redshifts of 8.66 and 8.72, are favoured for the second \dazle\ filter pair.

Due to the faintness of the sources that \dazle\ is designed to detect sensitivity is the overriding concern.  For this reason we have chosen to maximise sensitivity by matching the width of the \dazle\ filters to the expected line widths of the \ly\ emitters despite the fact that a wider filter would survey a wider redshift slice and hence a greater volume.  The specified FWHM of the \dazle\ filters is 10\AA, which corresponds to a velocity dispersion of approximately 300 km s$^{-1}$, towards the upper end of what might be expected for the width of a \lya\ line from a high redshift galaxy.  

The filter width is also constrained by the width of the OH windows.  The selected windows have widths of $\sim50$\AA\ which is five times the specified width of the filters, however the central wavelengths of a multilayer interference filters such as those used by \dazle\ depend on the angle of incidence of the light.  In \dazle\ this `phase effect' results in a shift to bluewards of the filter central wavelength as you move outwards from the centre of the field of view.  This wavelength shift amounts to -30\AA\ between the field centre and corners.  As a result filter bandpasses any greater than 10\AA\ would make it impossible to avoid the filter wings overlapping with an OH line for either small or large field angles.

Over the range 1-1.8\micron\ the total flux in OH lines is $\sim30$ times the total continuum emission.  A high level of off-band blocking is therefore required to prevent the OH lines from significantly adding to the total sky background.  An optical density (OD) of $>5$ has been specified which will ensure the off-band contribution is less than $1/4$ that of the $\sim10$\AA\ of continuum seen through the filter.

\begin{figure}[tbp]
\includegraphics[width=\textwidth]{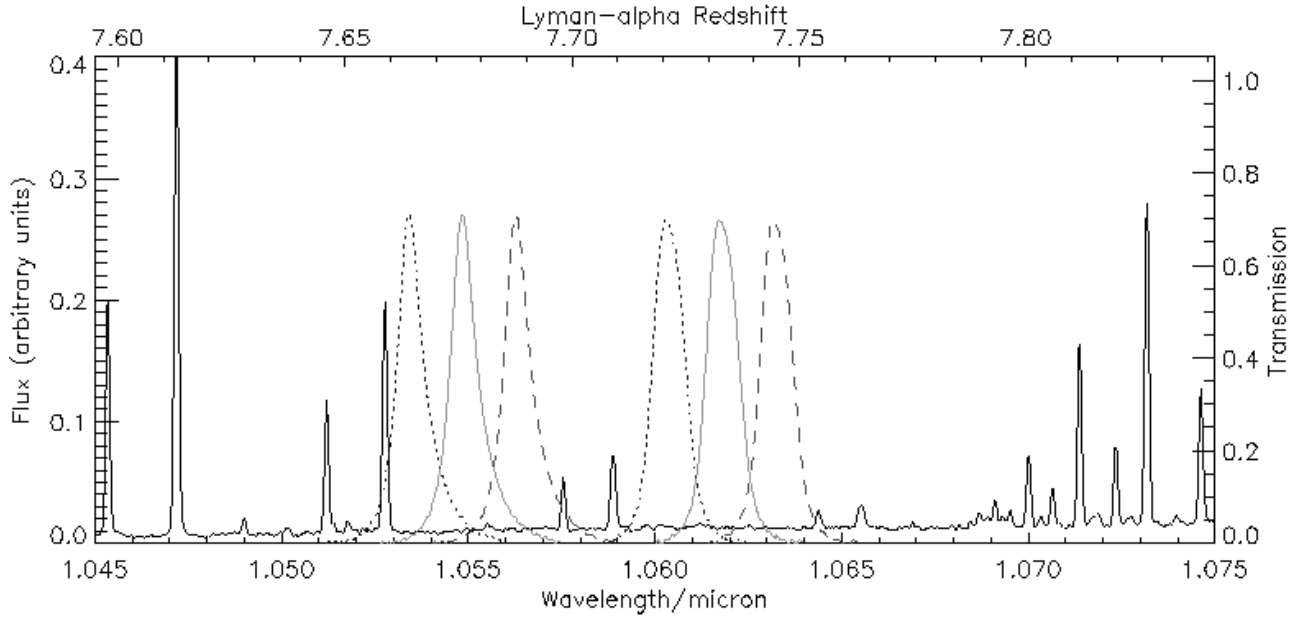}
\caption{\dazle\ narrowband filter profiles.  The dashed, solid and dotted lines are for the centre, edge and corner of the field of view respectively.  The Rousellot et al sky spectrum is also plotted.}
\label{fig:filters}
\end{figure}

\begin{table}[btp]
\centering
\caption[Narrowband filter properties]{Narrowband filter properties.\\}
\label{table:filters}
\begin{tabular}{|l|r|r|}
\hline
\rule[-1ex]{0pt}{3.5ex} Central wavelengh/\micron & 1.0563 & 1.0631 \\
\hline
\rule[-1ex]{0pt}{3.5ex} FWHM/\AA & 8 & 9 \\
\hline
\rule[-1ex]{0pt}{3.5ex} Peak transmission & 72\% & 72\% \\
\hline
\rule[-1ex]{0pt}{3.5ex} Blocking OD (0.5--1.5\micron) & $>5$ & $>5$ \\
\hline
\rule[-1ex]{0pt}{3.5ex} Clear diameter/mm & 110 & 110 \\
\hline
\end{tabular}
\end{table}

The \dazle\ filters represent a considerable manufacturing challenge.  The AAO entered into detailed discussions with Barr Associates Inc., the chosen manufacturer,  in order to arrive at a design which met our requirements while remaining practical.  This process included the reverse-engineering of the filter designs by the AAO to better understand the factors driving the complexity of the filters and the trade offs between the filters' different properties\cite{Cianci,newsletter}.  As result of this consulation process two options were offered by Barr, one design based on standard techniques and a second using an new filter fabrication technology currently under development.  The experimental technique was chosen as it promised higher transmission and reduced costs.  The final design employs 120mm diameter multilayer interference filters with a filter glass substrate.

The first pair of filters has been delivered by Barr together with extensive data on their performance.  Both meet the OD$>5$ blocking requirement, in fact for most of the wavelength range the OD exceeds 6.  The filter transmission profiles are plotted in Fig.\ \ref{fig:filters}.  To illustrate the phase effect the profiles are shown for the centre, edge and corner of the field of view.  Also plotted is the sky spectrum observed using ISAAC by Rousselot et al\cite{Rousselot_et_al}.  The key properties of these filters are listed in Table \ref{table:filters}.  The central wavelengths given in the table are for zero field angle.

\subsection{Optical Design}
\label{sect:design:optics}

\begin{figure}[tbp]
\includegraphics[width=\textwidth]{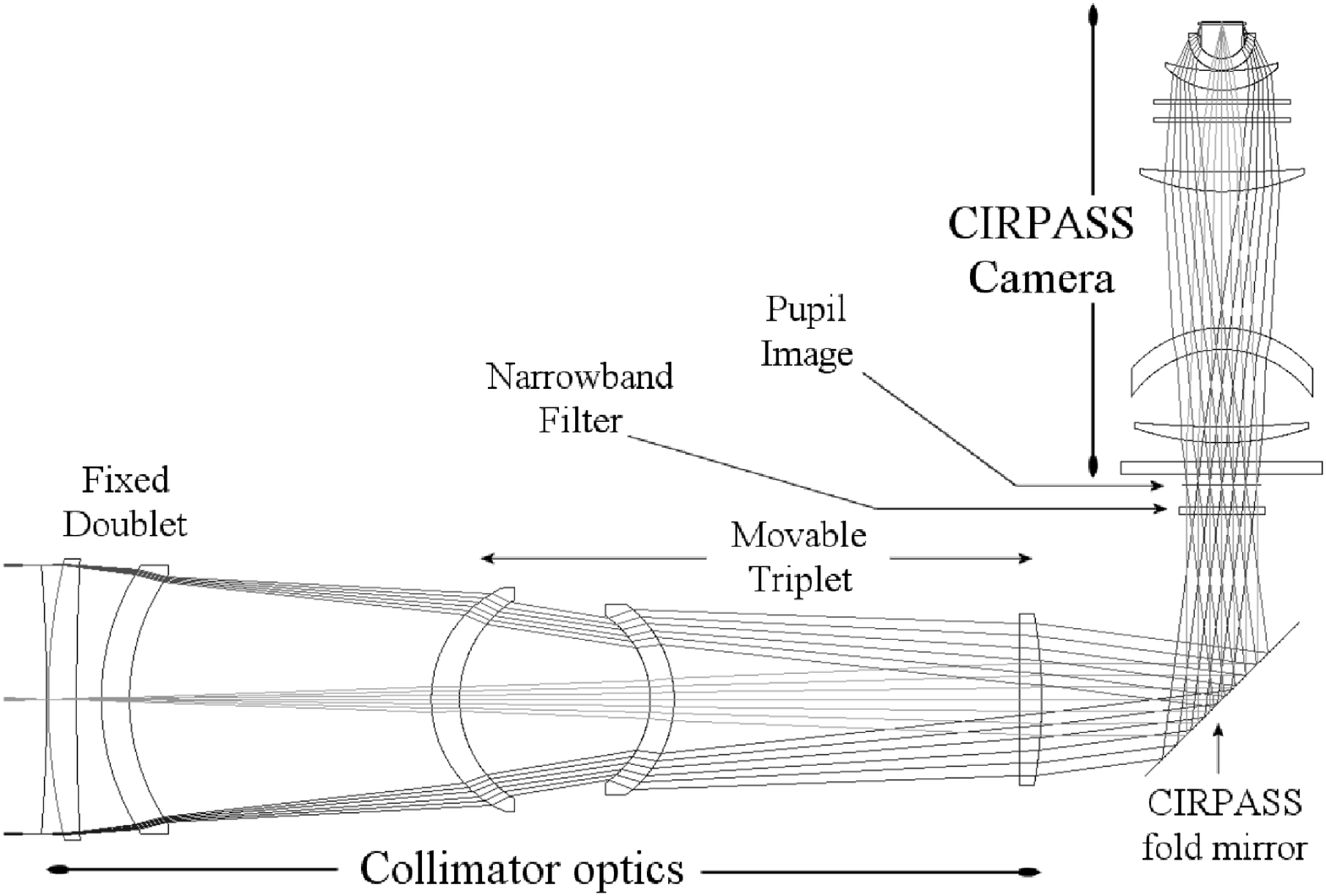}
\caption{Layout of the \dazle\ optical design.  The path of light rays from three different points in the field of view are shown.  The upper and lower ray bundle are the same distance from the centre ($4.83'$) as the corners of the field of view.  For scale, the distance from the front surface of the first lens to the sente of the fold mirror is 1.46m.}
\label{fig:optics}
\end{figure}

Fig.\ \ref{fig:optics} is a schematic of the \dazle\ optical design.  \dazle\ is a Nasmyth focus instrument and so the collimator accepts a horizontal beam from the telescope.  However, the CIRPASS camera must be mounted vertically and therefore a fold mirror is required between collimator and camera.  The collimated space contains a mask wheel (not shown), the narrowband filter wheel (one filter shown), and a cold stop at the pupil image.  Blocking filters are positioned in filter wheels between lenses 3 and 4 of the camera, and immediately in front of the detector.  Collimator and camera contain five elements each, and all lens surfaces are spherical.

\begin{figure}[t]
\centering
\includegraphics[width=0.6\textwidth]{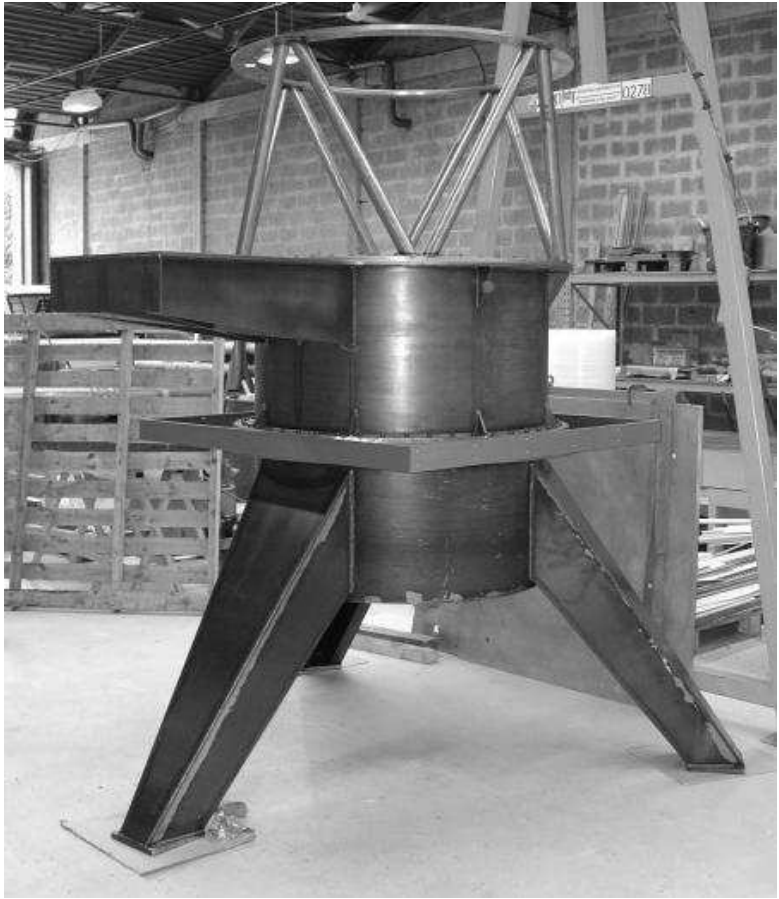}
\caption{The main structural elements of \dazle\ during trial assembly.  The overal height of the structure shown is 2.4 metres.}
\label{fig:structure}
\end{figure}

\begin{table}[bt]
\centering
\caption{Minimum ensquared energies.}
\label{table:ensquared}
\begin{tabular}{|l|r|r|r|r|r|}
\hline
\rule[-1ex]{0pt}{3.5ex} Wavelength (\micron) & 1.06 & 1.19 & 1.33 & 1.70 & Diff. limit\\
\hline
\rule[-1ex]{0pt}{3.5ex} 1 pixel & 74\% & 77\% & 76\% & 67\% & 96--94\% \\
\hline
\rule[-1ex]{0pt}{3.5ex} $2 \times 2$ pixels & 99\% & 99\% & 98\% & 94\% & 99.8--98\% \\
\hline
\end{tabular}
\end{table}

\dazle\ observations with a given filter pair are essentially monochromatic and so there is no need for achromatisation of the optics.  The change in operating wavelength when switching from one filter set to another will be accomodated by refocussing of the camera and collimator. The collimator is refocussed by movements of the rear lens triplet.  The front doublet is fixed in position as it will act as a `double-glazed' window for the instrument's refrigerated enclosure.  

The optical design produces images with a scale of $0.2''$ per pixel and $<1\%$ distortion.  The image quality is excellent throughout the J-band and remains good for H-band wavelengths.  Table \ref{table:ensquared} lists the minimum values of the ensquared energy over the field of view for wavelengths of 1.06, 1.19, 1.33 and 1.70\micron.  It also includes the diffraction limits for this range of wavelengths.  The ability to focus $\sim75\%$ of the incident energy within a single pixel's width ($0.2''$) for all J-band wavelengths means that \dazle's performance will always be seeing limited under anticipated VLT observing conditions.

The expected on-band optical throughput of \dazle\, taking into account all optical surfaces, blocking filters and narrowband filters, is 0.32.  The total efficiency of the \dazle -VLT combination, from photons entering the atmosphere to electrons at the detector, is estimated to be 0.13, which is comparable to other VLT near infrared imaging instruments such as ISAAC.

\subsection{Structural Design}
\label{sect:design:structure}

The optical axis at the VLT Nasmyth foci is over 2 metres above the Nasmyth platform. Consequently, the weight of the CIRPASS camera and \dazle\ optics must be supported at this considerable height by a fairly substantial structure.  Furthermore, as the VLT is situated in a region prone to earthquakes, the support structure must be stronger and more rigid that would be required just to support the weight of the instrument without distortion.  Using FEA the AAO have designed an welded steel structure which is extremely stiff, strong, and earthquake resistant.  The main elements of the structure are shown in Fig.\ \ref{fig:structure}.

\subsection{Field Rotation Compensation}
\label{sect:design:rotator}
\begin{figure}[t]
\centering
\includegraphics[width=0.8\textwidth]{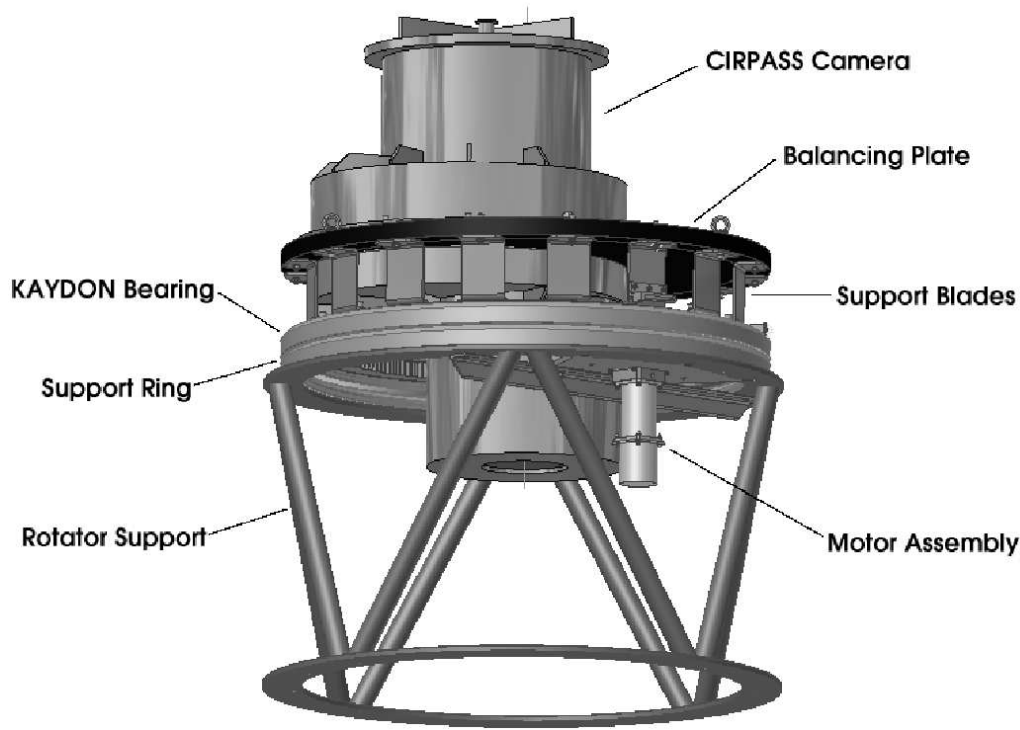}
\caption{The camera rotator assembly design.}
\label{fig:rotator}
\end{figure}

As, like almost all modern astronomical telescopes, the VLT has an altitude-azimuth mount its instruments must compensate for the apparent rotation of their fields of view.  The most common methods of achieving this are to either mount the instrument to an `adaptor-rotator' which physically rotates the entire instrument to match the field rotaton or to use optical de-rotators such as rotating dove prisms or `K-mirror' de-rotators.  Neither of these methods is entirely satisfactory for \dazle, however. Optical de-rotators are undesirable because they introduce additional optical surfaces, thereby reducing throughput, and \dazle's large field of view would require extremely large and costly de-rotator optics.  The adaptor-rotator cannot be used instead because of the need to keep the CIRPASS camera vertical.

The solution for \dazle\ is to rotate just the CIRPASS camera.  Rotating the camera about its axis compensates for field rotation without tilting the camera and requires no additional optical surfaces.  The rotator design is illustrated in Fig.\ \ref{fig:rotator}.  A closed loop control system using a magnetic positon encoder will provide the precision motion control required.

\section{EXPECTED CAPABILITIES}
\label{sect:expect}

\subsection{Sensitivity}
\label{sect:expect:sensitive}

\begin{figure}[tbp]
\centering
\includegraphics[width=\textwidth]{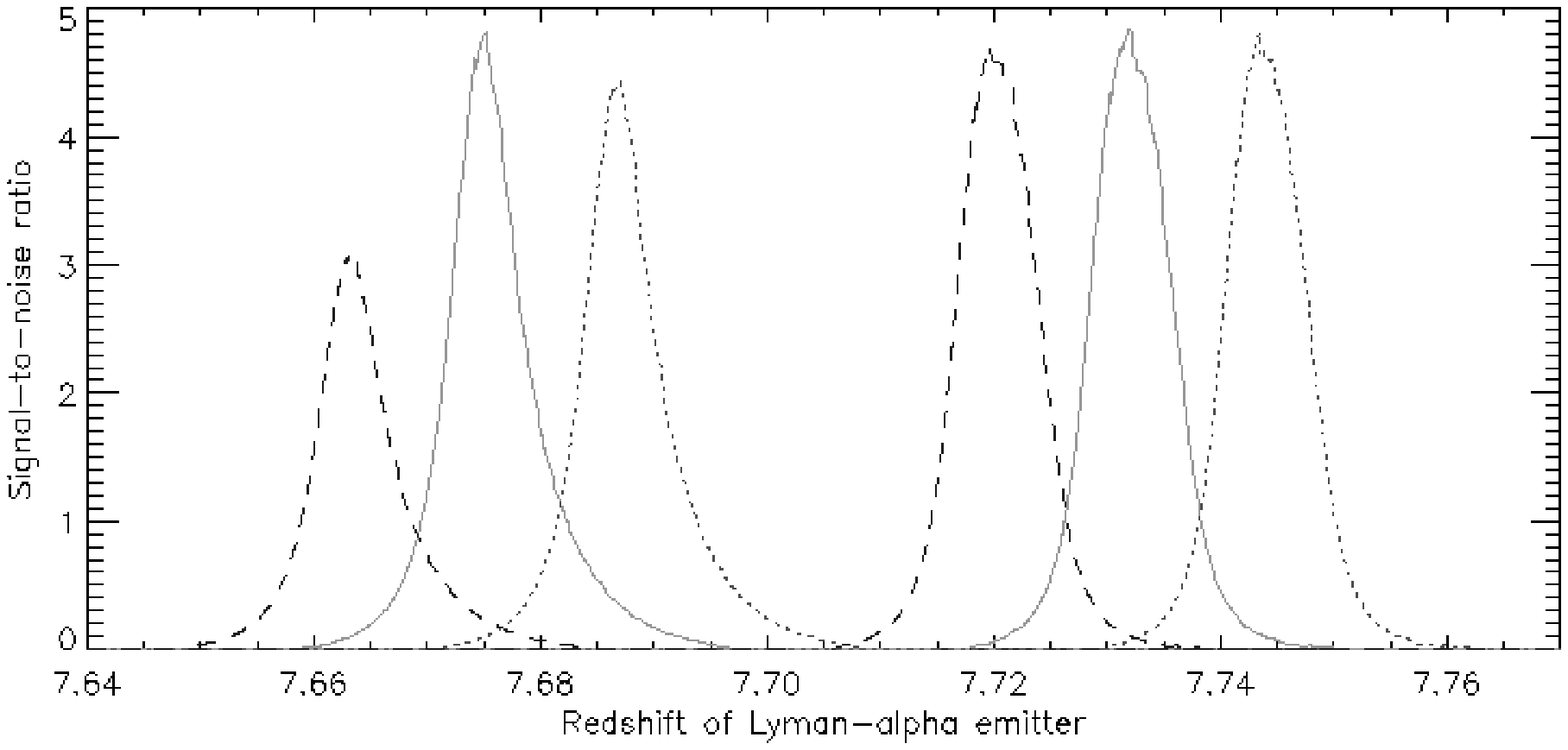}
\caption{Signal-to-noise in the summed images.  The dashed, solid and dotted lines are for the centre, edge and corner of the field of view respectively.  The total integration time is 36000s (15000s per filter) and the \lya\ flux is $2\times10^{-21}$Wm$^{-2}$.  The source velocity dispersion is 100km s$^{-1}$.}
\label{fig:s2n_100}
\end{figure}

\begin{figure}[tbp]
\centering
\includegraphics[width=\textwidth]{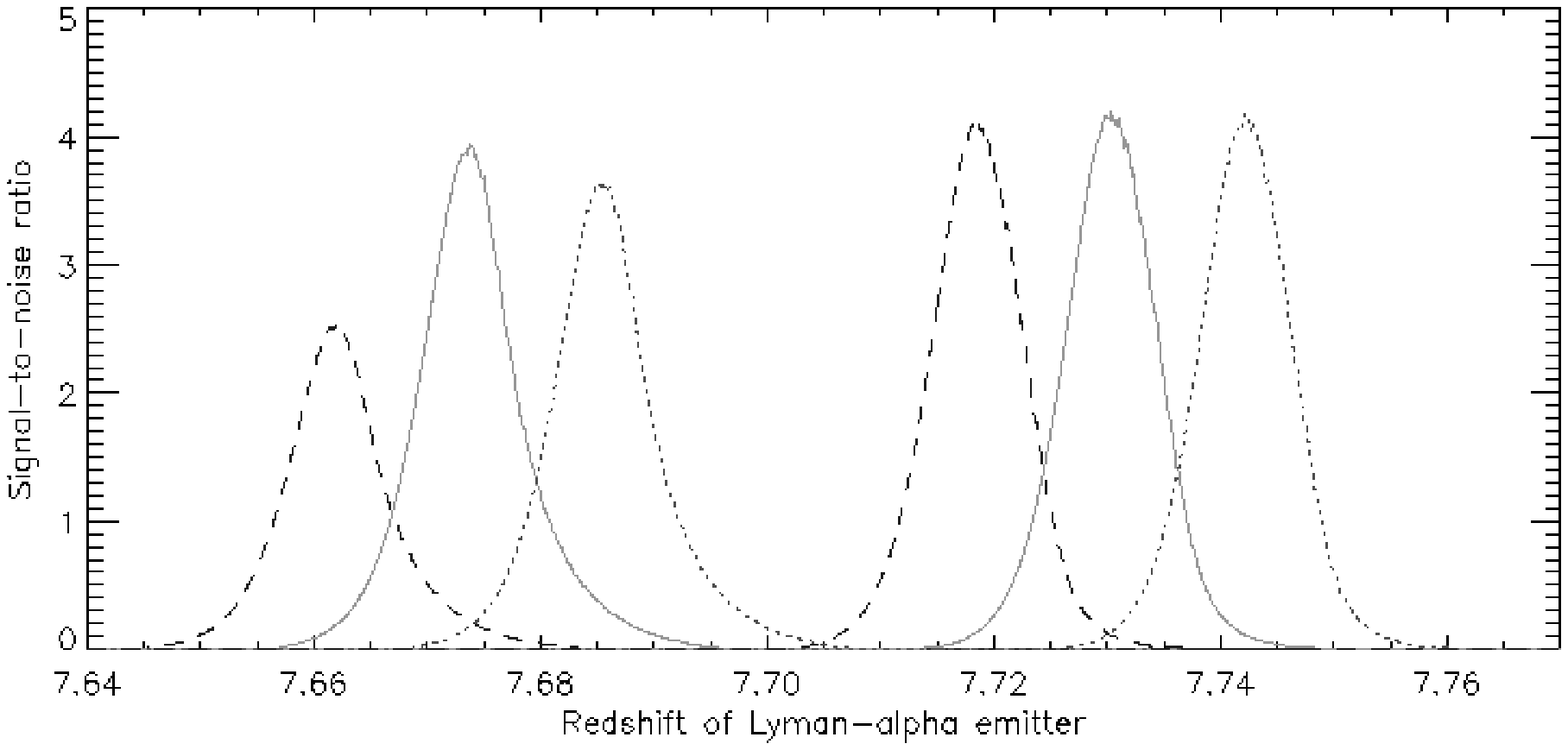}
\caption{Signal-to-noise in the summed images.  The dashed, solid and dotted lines are for the centre, edge and corner of the field of view respectively.  The total integration time is 36000s (15000s per filter) and the \lya\ flux is $2\times10^{-21}$Wm$^{-2}$.  The source velocity dispersion is 300km s$^{-1}$.}
\label{fig:s2n_300}
\end{figure}

The expected sensitivity of \dazle\ has been determined by calculating the signal-to-noise ratio as a function of redshift for a model \lya\ emitter.  The model \lya\ emitter spectrum used intrinsic \lya\ lines with Guassian profiles and FWHM velocity dispersions of 100 or 300 km s$^{-1}$. It was assumed that the blue half of the line is absorbed by the IGM, so that the observed \lya\ line appears as a `half-Gaussian'.   This is a highly simplified model but sufficient for sensitivity estimates.  See, for example, Santos 2004\cite{Santos2004} for detailed discussion of the likely observable properties of high redshift \lya\ emitters. 
To predict the sky background the model sky spectrum employed by the ESO exposure time calculators was used.  This model uses the Rousselot et al theoretical line list\cite{Rousselot_et_al} together with a continuum emission model and scales the intensities to match the observed broadband sky brightness at Paranal.  Detector dark current and read noise were also included in the calculations, but at all times the results were sky background limited.

The baseline observation sequence used was a total integration time of 36000 seconds, divided into 60 5 minute exposures for each of the two filters.  The key result is that \dazle\ will have a sensitivity of $\sim2\times10^{-21}$Wm$^{-2}$.  At this \ly\ line intensity the 36000s differenced image has a signal-to-noise ratio of $\sim3$, which would be considered a marginal detection.  This is all that's required of the differenced image, however, as once a possible line emitter has been located in the differenced image the corresponding position in the single filter images can be examined.  The single filter image containing the \ly\ signal will have a signal-to-noise approximately $\sqrt{2}$ times that of the differenced image as it contains the same signal but has noise contributions from only half the exposures.

Figs.\ \ref{fig:s2n_100} and \ref{fig:s2n_300} are plots of signal-to-noise ratio in the summed single filter images as a function of \lya\ emitter redshift.  Fig.\ \ref{fig:s2n_100} is for an intrinsic velocity dispersio of 100 km s$^{-1}$ while Fig.\ \ref{fig:s2n_300} is the result for 300 km s$^{-1}$.  The dotted, solid and dashed curves are for field angles of $0'$, $3.413'$ and $4.827'$ respectively, which correspond to the centre, edge and corners of the field of view.  The wavelength shift caused by the phase effect results in \dazle\ probing slightly different redshifts in different parts of its field of view.  The signal-to-noise in the summed images generally peaks at $\sim4$--5, which are reasonably robust detections.

\subsection{Volume Surveyed}
\label{sect:expect:volume}

The widths of the signal-to-noise peaks in Figs.\ \ref{fig:s2n_100} and \ref{fig:s2n_300} show the thicknesses of the `redshift slices' that the \dazle\ filters survey.  Defining the redshift slice as the interval over which the $5\sigma$ sensitivity limit is $\le5\times10^{-21}$Wm$^{-2}$ results in average widths of $\sim0.008$ for both filters.  Using the concordance cosmology ($H_0 = 70$km s$^{-1}$Mpc$^{-1}$, $\Omega_M = 0.30$, $\Omega_\Lambda = 0.7$) this redshift interval converts to a line of sight comoving distance of $\sim2.4$Mpc at $z\sim7.70$.  At the same redshift the \dazle\ field of view, $6.83'\times6.83'$, equates to 17.6Mpc$\times$17.6Mpc.  The total comoving volume surveyed by the filter pair in a single pointing is approximately 1600Mpc$^3$.

\subsection{Number Counts}
\label{sect:expect:numbers}

There are many poorly constrained parameters which determine the observed luminosity of a high redshift \lya\ emitter.  These include the star formation rate, the escape fraction of ionising photons, dust, the velocity dispersion, galactic winds, the size of the HII region surrounding the galaxy and the ionisation state of the intergalactic medium.  The influence of all these factors makes it difficult to relate the \dazle\ sensitivity to particular properties of high redshift galaxies such as a given star formation rate, however the sensitivity limit should correspond approximately to star formation rates of a few solar masses per year\cite{Santos2004}.  The lack of firm constraints on the \lya\ luminosity function at the \dazle\ target redshifts makes predictions of the expected numbers of detections difficult.  Predicted $z>7$ \lya\ luminosity functions in the literature vary widely. For example, the modelling of Thommes and Heisenheimer predicts 0.2--10 or 0.04--8 \lya\ emitters brighter than $2\times10^{-21}$Wm$^{-2}$ per 1600Mpc$^3$ at redshifts of 6.6 and 9.3 respectively\cite{Thommes2004}.  In contrast, Barton et al predict 140 \lya\ emitters per 1600Mpc$^3$ at $z=8.227$ for the same flux limit\cite{Barton2004}.  Stiavelli, Fall and Panagia, on the other hand, predict $\sim1$ \lya\ emitter  per 1600Mpc$^3$ brighter than $2\times10^{-21}$Wm$^{-2}$ at $z>6$\cite{Stiavelli2004}.  Given these predictions 1--10 detectable \lya\ emitters per \dazle\ field does not seem unreasonable, however this figure is very uncertain. This level of uncertainty reflects, in part, how much we can learn from high redshift \lya\ surveys.

\section{SUMMARY AND TIMELINE}
\label{sect:summary}

\dazle\ is a special purpose near infrared narrowband differential imager with the sole purpose of detecting redshifted \lya\ emission from star forming galaxies at $z>7$.  It is being built by the Institute of Astronomy, Cambridge, in collaboration with the Anglo-Australian Observatory, and is intended to be a visitor instrument for the ESO VLT.  \dazle\ will use high resolution ($R=1000$) filters in order to exploit low background windows in the sky emission spectrum.  This will enable \dazle\ to reach sensitivities of $\sim2\times10^{-21}$Wm$^{-2}$ in 10 hours .  This impressive sensitivity will enable \dazle\ to detect $z>7$ galaxies with star formation rates as low as a few solar masses per year.  With each pointing \dazle\ will survey a comoving volume of $\sim1600$Mpc$^3$.  Published predictions for $z>7$ \lya\ emitter abundances and luminosities vary widely, however it is not unreasonble to expect \dazle\ to detect of order 1--10 \lya\ emitters per pointing.

At the time of writing \dazle\ is well on the way to completion.  All major components, such as the support structure, optics and filters have already been manufactured.  The cryogenic camera to be used within \dazle\ is already tried and tested, having been used on numerous occasions in the CIRPASS spectrograph.  Integration and testing are now in progress at the IoA, and the instrument is expected to be completed by Dec 2004.  It is hoped that commissioning will take place in Q2-Q3 2005, with a program of science observations of the GOODS-Chandra Deep Field South taking place in Q3-Q4 2005.  We plan to acquire the second narrowband filter pair in 2006 while continuing the science program.

\appendix    

\acknowledgments     
The authors warmly thank the Raymond and Beverly Sackler Foundation and PPARC for funding this project.  We would also like to thank all the other members of the AAO that have worked on the \dazle\ project for their valuable contributions; John O'Byrne, Vlad Churilov, Chris Evans, Roger Haynes, Mark Hilliard, Andrew McGrath and Stan Mizarski.  Thanks are also due to Craig Mackay and Jim Pritchard of the IoA for their assistance.


\bibliography{../thesis}   

\begin{thebibliography}{10}

\bibitem{Hu2002}
E.~M. {Hu}, L.~L. {Cowie}, R.~G. {McMahon}, P.~{Capak}, F.~{Iwamuro}, J.-P.
  {Kneib}, T.~{Maihara}, and K.~{Motohara}, ``{A Redshift z=6.56 Galaxy behind
  the Cluster Abell 370},'' {\em ApJ} {\bf 568}, pp.~L75--L79, Apr. 2002.

\bibitem{Kodaira2003}
K.~{Kodaira}, Y.~{Taniguchi}, N.~{Kashikawa}, N.~{Kaifu}, H.~{Ando},
  H.~{Karoji}, M.~{Ajiki}, M.~{Akiyama}, K.~{Aoki}, M.~{Doi}, S.~S. {Fujita},
  H.~{Furusawa}, T.~{Hayashino}, M.~{Imanishi}, F.~{Iwamuro}, M.~{Iye}, K.~S.
  {Kawabata}, N.~{Kobayashi}, T.~{Kodama}, Y.~{Komiyama}, G.~{Kosugi},
  Y.~{Matsuda}, S.~{Miyazaki}, Y.~{Mizumoto}, K.~{Motohara}, T.~{Murayama},
  T.~{Nagao}, K.~{Nariai}, K.~{Ohta}, Y.~{Ohyama}, S.~{Okamura}, M.~{Ouchi},
  T.~{Sasaki}, K.~{Sekiguchi}, K.~{Shimasaku}, Y.~{Shioya}, T.~{Takata},
  H.~{Tamura}, H.~{Terada}, M.~{Umemura}, T.~{Usuda}, M.~{Yagi}, T.~{Yamada},
  N.~{Yasuda}, and M.~{Yoshida}, ``{The Discovery of Two Lyman {$\alpha$}
  Emitters beyond Redshift 6 in the Subaru Deep Field},'' {\em PASJ} {\bf 55},
  pp.~L17--L21, Apr. 2003.

\bibitem{WMAP2003a}
A.~{Kogut}, D.~N. {Spergel}, C.~{Barnes}, C.~L. {Bennett}, M.~{Halpern},
  G.~{Hinshaw}, N.~{Jarosik}, M.~{Limon}, S.~S. {Meyer}, L.~{Page}, G.~S.
  {Tucker}, E.~{Wollack}, and E.~L. {Wright}, ``{First-Year Wilkinson Microwave
  Anisotropy Probe (WMAP) Observations: Temperature-Polarization
  Correlation},'' {\em ApJS} {\bf 148}, pp.~161--173, Sept. 2003.

\bibitem{dietrich2003}
M.~{Dietrich}, F.~{Hamann}, I.~{Appenzeller}, and M.~{Vestergaard}, ``{Fe II/Mg
  II Emission-Line Ratio in High-Redshift Quasars},'' {\em ApJ} {\bf 596},
  pp.~817--829, Oct. 2003.

\bibitem{corbin2003}
M.~{Corbin}, W.~{Freudling}, and K.~{Korista}, ``{Iron Emission in $z\approx6$
  QSOs and its Possible Implications},'' in {\em {AGN Physics with the Sloan
  Digital Sky Survey}},  G.~{Richards} and P.~{Hall}, eds., {\em ASP Conference
  Series} {\bf 311}, July 2003.

\bibitem{maiolino2003}
R.~{Maiolino}, Y.~{Juarez}, R.~{Mujica}, N.~M. {Nagar}, and E.~{Oliva},
  ``{Early Star Formation Traced by the Highest Redshift Quasars},'' {\em ApJ}
  {\bf 596}, pp.~L155--L158, Oct. 2003.

\bibitem{barth2003}
A.~J. {Barth}, P.~{Martini}, C.~H. {Nelson}, and L.~C. {Ho}, ``{Iron Emission
  in the z = 6.4 Quasar SDSS J114816.64+525150.3},'' {\em ApJ} {\bf 594},
  pp.~L95--L98, Sept. 2003.

\bibitem{ajd}
A.~J. {Dean}, {\em {CIRPASS: The Cambridge Infrared Panoramic Survey
  Spectrograph}}.
\newblock PhD thesis, University of Cambridge, July 2002.

\bibitem{Rousselot_et_al}
P.~{Rousselot}, C.~{Lidman}, J.-G. {Cuby}, G.~{Moreels}, and G.~{Monnet},
  ``{Night-sky spectral atlas of OH emission lines in the near-infrared},''
  {\em A\&A} {\bf 354}, pp.~1134--1150, Feb. 2000.

\bibitem{Abrams}
M.~C. {Abrams}, S.~P. {Davis}, M.~L.~P. {Rao}, R.~J. {Engleman}, and J.~W.
  {Brault}, ``{High-resolution Fourier transform spectroscopy of the Meinel
  system of OH},'' {\em ApJS} {\bf 93}, pp.~351--395, July 1994.

\bibitem{Mies}
F.~{Mies}, ``{Calculated vibrational transition probabilities of OH
  ($X^{2}\Pi$)},'' {\em J.Mol.Spec} {\bf 118}(507), 1974.

\bibitem{Maihara}
T.~{Maihara}, F.~{Iwamuro}, T.~{Yamashita}, D.~N.~B. {Hall}, L.~L. {Cowie},
  A.~T. {Tokunaga}, and A.~{Pickles}, ``{Observations of the OH airglow
  emission},'' {\em PASP} {\bf 105}, pp.~940--944, Sept. 1993.

\bibitem{Lord}
S.~D. {Lord}, ``{A New Software Tool for Computing Earth's Atmospheric
  Transmission of Near- and Far-Infrared Radiation},'' Dec. 1992.
\newblock NASA Technical Memor. 103957.

\bibitem{Cianci}
S.~{Cianci}.
\newblock PhD thesis, University of Sydney, 2002.

\bibitem{newsletter}
S.~{Cianci}, ``{Reverse-Engineering Filter Designs for \dazle}.'' In AAO
  Newsletter No.\ 103, Nov. 2003.

\bibitem{Santos2004}
M.~R. {Santos}, ``{Probing reionization with Lyman {$\alpha$} emission
  lines},'' {\em MNRAS} {\bf 349}, pp.~1137--1152, Apr. 2004.

\bibitem{Thommes2004}
E.~{Thommes} and K.~{Mesienheimer}, ``{The expected abundance of \lya\ emitting
  primeval galaxies. I. General model predictions.},'' {\em pre-print} , 2004.
\newblock astro-ph/0312363.

\bibitem{Barton2004}
E.~J. {Barton}, R.~{Dav{\' e}}, J.~T. {Smith}, C.~{Papovich}, L.~{Hernquist},
  and V.~{Springel}, ``{Searching for Star Formation beyond Reionization},''
  {\em ApJ} {\bf 604}, pp.~L1--L4, Mar. 2004.

\bibitem{Stiavelli2004}
M.~{Stiavelli}, S.~M. {Fall}, and N.~{Panagia}, ``{Observable Properties of
  Cosmological Reionization Sources},'' {\em ApJ} {\bf 600}, pp.~508--519, Jan.
  2004.

\end{thebibliography}
\bibliographystyle{spiebib}   

\end{document}